\documentclass[journal]{Figures/IEEEtran}

\usepackage{amsmath}
\usepackage{amssymb}
\usepackage{color}
\usepackage{dsfont}
\usepackage[numbers]{natbib}
\usepackage{savesym}
\usepackage{graphicx}
\usepackage{psfrag,graphicx,epsfig}
\usepackage{algorithm,algorithmic,xspace}
\usepackage{subfig}
\usepackage{float}
\usepackage{graphicx}
\usepackage{epstopdf}
\usepackage{url}

\newcommand{\nnum}{\nonumber}

\newcommand{\real}{\mathds{R}}

\newcommand{\CC}{\mathcal{C}}

\newcommand{\EE}{\mathcal{E}}
\newcommand{\FF}{\mathcal{F}}
\newcommand{\GG}{\mathcal{G}}

\newcommand{\PP}{\mathcal{P}}

\newcommand{\VV}{\mathcal{V}}

\newcommand{\Ss}{\mathcal{S}}
\newcommand{\RR}{\mathcal{R}}

\DeclareMathOperator*{\argmax}{arg\,max}

\newtheorem{theorem}{\bf Theorem}[section]

\newtheorem{definition}{\bf Definition}[section]
\newtheorem{remark}{\bf Remark}[section]

\newtheorem{claim}{\bf Claim}
\newtheorem{assumption}{\bf Assumption}[section]

\begin{document}
\setlength{\abovedisplayskip}{5pt} % remove align margin
\setlength{\belowdisplayskip}{5pt}
\setlength{\abovedisplayshortskip}{0pt}
\setlength{\belowdisplayshortskip}{0pt}

\let\WriteBookmarks\relax
\def\floatpagepagefraction{1}
\def\textpagefraction{.001}

\title{Optimal Bi-level Lottery Design for Multi-agent Systems}

%  \tnotemark[1]

%  \tnotetext[1]{This document is the results of the research project funded by the National Science Foundation.}

\author{
Hunmin Kim, and Minghui Zhu\thanks{Hunmin Kim (hunminkim3@gmail.com) is with the Department of Mechanical Science and Engineering, University of Illinois, 605 East Springfield Avenue, Champaign, IL 61820.}
\thanks{Minghui Zhu (muz16@psu.edu) are with the School of Electrical Engineering and Computer Science, Pennsylvania State University, 201 Old Main, University Park, PA 16802.}
}

\onecolumn

This work has been submitted to the IEEE for possible publication. Copyright may be transferred without notice, after which this version may no longer be accessible.

\twocolumn

\newpage

\maketitle

\begin{abstract}
Entities in multi-agent systems may seek conflicting subobjectives, and this leads to competition between them. To address performance degradation due to competition,
we consider a bi-level lottery where a social planner at the high level selects a reward first and, sequentially, a set of players at the low level jointly determine a Nash equilibrium given the reward. The social planner is faced with efficiency losses where a Nash equilibrium of the lottery game may not coincide with the social optimum. We propose an optimal bi-level lottery design problem as finding the least reward and perturbations such that the induced Nash equilibrium produces the socially optimal payoff. We formally characterize the price of anarchy and the behavior of public goods and Nash equilibrium with respect to the reward and perturbations. We relax the optimal bi-level lottery design problem via a convex approximation and identify mild sufficient conditions under which the approximation is exact.
\end{abstract}
% \begin{keywords}   Multi-agent systems, Mechanism design, and Game theory  \end{keywords}

\section{Introduction}
Advanced information and communications technologies have been stimulating the rapid emergence of multi-agent systems where many spatially distributed agents interact with each other to accomplish complex missions. Substantial effort has been spent on analysis, design, and control of multi-agent systems~\cite{bertsekas1989parallel,bullo2009distributed,mesbahi2010graph,ren2008distributed,zhu2015distributed}.
In many practical scenarios, agents are non-cooperative and seek for heterogeneous (or even conflicting) subobjectives. This leads to competition over limited resources and degradation of system-wide performance.
Common practices to address the issue include mechanism design or incentive design, which modify agents' preferences via side payment or pricing so that individual interests are aligned with social welfare.

Mechanism design has been studied when the designer has complete and incomplete information on agents' types \cite{shen2007optimal}.
There are several classes of mechanism design with incomplete information. An auction consists of multiple bidders who submit bids according to their valuations of items being auctioned, and an auctioneer who sequentially determines item allocation and pricing. The most well-known auction mechanism is Vickrey-Clarke-Groves (VCG) auction~\cite{EC:71,TG:73,WV:61} which is efficient and incentive compatible~\cite{green1977characterization,holmstrom1979groves}.
Contract theory~\cite{bolton2005contract} studies how the principal constructs a contract in the presence of asymmetric information of the agent(s). There are three major models; i.e., moral hazard~\cite{shavell1979risk} (the agents have hidden information after the contract), adverse selection~\cite{baron1982regulating} (the agents have hidden information before the contract), and signaling~\cite{spence1973job} (the agents provide some confidential information).
Mechanisms have been applied to various areas, including smart grid~\cite{ma2014incentive}, communication networks~\cite{huang2006auction,Johari.Tsitsiklis:04}, and transportation networks~\cite{DM-BP-NR:09}.
Moreover, mechanism design has been extended to dynamic scenarios where agents are incentivized to follow specified algorithms and solve computation or control problems~\cite{NN-AR:99,AP-BF-DCP:06,TT-FF-CL:13}.

%Moreover, incentive design has been extended to dynamic scenarios and is comprised of two classes: directed and indirected. Algorithmic mechanism design is directed where agents are incentivized to follow specified algorithms and solve computation or control problems~\cite{NN-AR:99,AP-BF-DCP:06,TT-FF-CL:13}.

In mechanism design with complete information, the designer does not experience a lack of knowledge of agents' types. Therefore, the designer can deduce the agents' responses to its policy choice. This class includes optimal taxation~\cite{mirrlees1971exploration}, game design~\cite{arslan2007autonomous,marden2009cooperative}, incentive control~\cite{TB:84,YH-PL-GO:82}, and lottery~\cite{JM:00}. In optimal taxation, agents maximize their own utility functions by choosing labor time and consumption, and the designer chooses the optimal tax function to maximize the overall utility.
In game design, the designer chooses the utility functions of agents to achieve specific control objectives, and the agents maximize their utilities. Incentive control is an incentive design in a dynamic environment where agents' choices are affected by rewards or prices such that social welfare can be optimized.

As one kind of mechanism design with complete information, fixed prize lotteries have been applied to several field experiments and proven to stimulate agents' or players' investments effectively.
INSINC project in Singapore~\cite{pluntke2013insinc} is an ongoing real-world implementation of a lottery scheme for commuters who use public transportation to travel off-peak hours.
The lottery scheme successfully reduces around $7.5\%$ of peak time demand.
A similar project named INSTANT~\cite{DM-BP-NR:09} is conducted in India and results in more than $20\%$ of commuter shifts.
Research~\cite{barbieri2015incentives} uses the boarding passes of local public transportation as lottery tickets, showing that the lottery increases the provision of public goods and reduces free riders. In~\cite{laguilles2011can,robertson2005response,sundar1992mail}, experiments are conducted to show that lottery based incentives can effectively increase survey response rates.
Moreover, lottery based incentives have been used in demand response in the smart grid~\cite{li2015energy,schwartz2014demand}, mobile crowdsensing for
traffic congestion and air pollution~\cite{djehiche2016mean}, and Internet congestion~\cite{loiseau2011congestion}.

Substantial effort has been exerted to develop the fundamental theory of lotteries.
Seminal paper~\cite{JM:00} studies that fixed prize lotteries alleviate the free-rider problem and nudge higher levels of public good provisions as well as aggregate payoff than voluntary contributions. A larger reward results in a greater public good and aggregate payoff. The results have been extended by many researchers. In~\cite{lange2007using}, a multi-prize lottery is studied considering risk preferences; i.e., risk-neutral versus risk-averse.
A sequential lottery is investigated in~\cite{damianov2017disclosure} in which it can sell more tickets than a one-level lottery.
Paper~\cite{pecorino2007lotteries} analyzes public good on player size, and extends the results to a rival public good case; i.e., players benefit from a portion of public goods.

%A new two-step lottery where a reward is funded voluntarily in the first step, outperforms a voluntary contribution~\cite{lange2006providing}.

\textbf{Contributions.}
In the classic lottery schemes, the competition among the players induces efficiency losses; i.e., a Nash equilibrium of a lottery game may not coincide with its social optimum. The social optimum is only achieved when an infinite reward is given~\cite{JM:00}. To address the issue, we introduce perturbation parameter chosen by the social planner and formulate an optimal bi-level lottery design problem where a Nash equilibrium of a lottery game induces a socially optimal payoff with the least reward and perturbations.
On top of this, we impose general convex inequality constraints to encompass physical constraints and social planner's interest.
We analyze the properties of low-level Nash equilibrium, including the price of anarchy as well as the behavior of public goods and Nash equilibrium with respect to the reward and perturbations.
By leveraging the above analytical results, we derive a convex approximation of the optimal bi-level lottery design problem and identify mild sufficient conditions under which the approximation is exact. Our results are verified via a case study on demand response in the smart grid.

This paper is enriched from preliminary version~\cite{kim2015optimal}, and includes a set of new results. In particular, this paper derives new properties of Nash equilibrium and public goods as well as more practical bounds on price of anarchy. Additionally, this paper introduces a convex approximation of the optimal lottery design problem and show that there is no approximation error. Further, a case study on demand response is provided to demonstrate the developed results.

\textbf{Paper organization.}
In Section~\ref{lin.sec:Problem}, we discuss a classic bi-level lottery scheme and its limitations. To alleviate the fundamental limitation of efficiency losses, we introduce a new perturbed bi-level lottery model and formulate the optimal bi-level lottery design problem in Section~\ref{lin.sec:s_mechanism}.
In Section~\ref{lin.sec:an}, we analyze the properties of low-level Nash equilibrium.
Based on the properties, we relax the optimal bi-level lottery design problem as a convex optimization problem in Section~\ref{leastP}. Section~\ref{sec:case} presents a case study on demand response.

\section{Preliminaries}
\label{lin.sec:Problem}
We introduce a classic bi-level lottery scheme in~\cite{JM:00} and outline its procedure in Section~\ref{sec:preliminary} to~\ref{sec:High}. Section~\ref{sec:problem_form} discusses its limitations and motivates our problem. Please refer to~\cite{JM:00} for comprehensive discussions.

\subsection{Payoff model}\label{sec:preliminary}
Consider a social planner who holds a lottery and a set of players $\VV \triangleq \{1,2,\cdots,N\}$ who participate in the lottery.
Before holding the lottery, the social planner announces a public good to be financed by the net profit of the lottery.
The public good provision consists of the net profit of the lottery and benefits all the players, but the amount of benefits may be different.
Each player chooses a benefit function corresponding to the announced public good.
Then, the social planner chooses a reward $R$ from an action set $\RR = (0,\infty)$.
%Given reward $R$, each player $i$ invests $s_i$ to the lottery from an action set $\Ss_i = [0, w_i]$ where $w_i$ denotes the amount of investable wealth of player $i$, and ${\Ss} \triangleq {\Ss}_1 \times \cdots \times {\Ss}_N$ denotes the joint action set. A player pays investment $-s_i$ and receives a portion of the reward $R$ proportional to investment over the total investment, and also benefits from the net profit.
Given reward $R$, each player $i$ invests $s_i$ to the lottery from an action set $\Ss_i = [0, w_i]$, and receives a portion of the reward $R$ proportional to its own investment over the total investment, and also gets benefits from the public good, where $w_i$ denotes the amount of investable wealth of player $i$.
The action profile $s \triangleq \{s_i\}_{i \in \VV} \in {\Ss}$ can be expressed as $\{s_i,s_{-i}\}$ where $s_{-i}$ denotes the action profile other than player $i$; i.e., $s_{-i} \triangleq \{s_j\}_{j \in \VV \setminus \{i\}}$.
Given reward $R$, payoff function $u_i: {\Ss} \rightarrow {\real}$ associated with $i$ is described by:
\begin{align}
u_i(s,R) \triangleq
\left\lbrace
\begin{array}{cc}
\frac{s_i}{\bar{s}}R + h_i(\bar{s}-R) -s_i, &{\rm for } \ \bar{s} \geq R\\
0, & {\rm otherwise} \\
\end{array}
\right.
\label{e140}
\end{align}
where $\bar{s} \triangleq \sum_{i \in \VV} s_i$.
Payoff function~\eqref{e140} indicates that the lottery holds only when total investment $\bar{s}$ exceeds or equals to reward $R$; otherwise, the social planner cancels the lottery and returns the investments to the players.
The first term $\frac{s_i}{\bar{s}}R$ represents the portion of reward from the lottery and 
the rate $\frac{s_i}{\bar{s}}$ can be seen as the probability of winning if a raffle gives the reward, and the players are risk-neutral; i.e., they consider the expected reward $\frac{s_i}{\bar{s}}R$ as the utility.
The last term $-s_i$ denotes the cost of player $i$.
The marginal benefit function $h_i: \real_{\geq 0}\rightarrow\real_{\geq 0}$ represents any benefit obtained from the pre-announced public good and is a function of net profit $\bar{s}-R$.
It can also be seen as the agent's private valuation on excess utility, which stimulates its investment.

In the classic lottery, there are two important assumptions. The first one is that players experience diminishing marginal utility from the provision of the public good, which is a classic assumption in social economics~\cite{veenhoven1991happiness,frey2010happiness}.
The other assumption is that the public good is socially desirable; i.e., 
financing non-zero public good increases network-wide payoffs.
The formalization of these assumptions is as:
\begin{assumption}
Function $h_i$ is twice differentiable, strictly increasing, strictly concave, $h_i(0) = 0$, $\sum_{i \in \VV}\frac{\partial h_i(0)}{\partial v} > 1$, and $\lim_{v \rightarrow \infty} \frac{\partial h_i (v)}{\partial v} =0$.
\label{nasm1}
\end{assumption}
In applications, $h_i$ has been seen as any side benefit which is generated by the net-profit (or social benefits). For instance, in demand response, $h_i$ denotes the level of (inverse) harzard~\cite{li2015energy} and any side payment from the net-profit~\cite{schwartz2014demand}. In internet congestion, $h_i(\cdot)=-L(\cdot)$ where $L(\cdot)$ is a disutility due to congestion~\cite{loiseau2011incentivej}.

%\subsection{Motivating example}

\subsection{Low-level decision making - Nash equilibrium}
Given $R$ and $s_{-i}$, player $i$ chooses $s_i$ to maximize its own payoff as follows:
\begin{align*}
\max_{s_i \in {\Ss}_i} u_i(s,R).
\end{align*}
The collection of local optimization problems induces a non-cooperative game among the players and the game is parameterized by $R$.
Nash equilibrium~\cite{nash1951non} defines the solution of the game.
\begin{definition} Given $R$,
the action profile $s^*(R)$ is a (pure) Nash equilibrium if $u_i(s_i',s_{-i}^*(R),R)\leq u_i(s^*(R),R)$ for $\forall s_i'$ $\in {\Ss}_i, \forall i \in \VV$.\label{def1}
\end{definition}
Note that Nash equilibrium $s^*(R)$ highlights its dependency on reward $R$.

\subsection{High-level decision making - Social optimum}\label{sec:High}
The lottery is a bi-level decision making (or a hierarchical optimization) problem where the social planner at the high level selects reward first and, sequentially, the players at the low level jointly determine a Nash equilibrium given the reward. The social planner aims to choose reward $R$ to maximize the aggregate payoff of the players at the induced Nash equilibrium:
\begin{align}
\begin{array}{c}
\underset{R \in {\RR}}{\max}
\underset{i \in \VV}{\sum} u_i(s^*(R),R)\\
{\rm s.t.}\quad g(s^*(R),R) \leq 0
\end{array}
= 
\begin{array}{c}
\underset{R \in {\RR}}{\max} \underset{i \in \VV}{\sum}h_i(G(R))-G(R)\\
{\rm s.t.}\quad g(s^*(R),R) \leq 0
\end{array}
%&\max_{R \in {\RR}}\sum_{i \in \VV} u_i(s^*(R),R) = \max_{R \in {\RR}} \sum_{i \in \VV}h_i(G(R))-G(R)\nnum\\
%&\ {\rm s.t.}\quad g(s^*(R,c),R,c) \leq 0,
\label{eq009}
\end{align}
where $G(R) \triangleq \bar{s}^*(R)-R$ is referred to as the \emph{public good} which is obtained by transforming the net profit $\bar{s}^*(R)-R$ into $G(R)$, on a one-for-one basis.
The hierarchical nature of the problem requires the social planner to predict the low-level Nash equilibrium when making decisions at the high level.

Inequality constraint $g(s^*(R),R) \leq 0$ expresses physical constraints (e.g., safety constraint, and flow capacity) and social planner's interest (e.g., the amount of required investments) as shown in Section~\ref{sec:case}
where $g: {\Ss} \times {\RR} \rightarrow {\real}^{m}$ is a vector of convex functions $g_{\ell}(s^*(R),R)$ for $\ell=1,2,\cdots,m$.
The convex inequality constraint $g(s^*(R),R)$ $\leq 0$ is absent in the classic bi-level lottery in~\cite{JM:00}.
\begin{assumption}
Function $g_{\ell}(s^*,R)$ is convex with respect to its arguments $s^*$ and $R$ for $\ell=1,2,\cdots,m$.
\label{nasm3}
\end{assumption}

%The lottery design is an incentive design where messages are being exchanged between the social planner and agents. The agents provide the marginal benefit function $h_i$, and then the social planner provides $(R,c)$.

% The lottery design is an incentive design where the designer does not experience a lack of information.

The lottery design is an incentive design with complete information. Given the marginal benefit function $h_i$, the social planner provides $(R,c)$.

\subsection{Limitations}\label{sec:problem_form}
When the constraint $g(s^*(R),R) \leq 0$ is absent, Assumption~\ref{nasm1} ensures that there exists a unique socially optimal public good (Proposition 2.1 in~\cite{JM:00})
\begin{align*}
G^* = \argmax_{G \in [0,\infty)}\sum_{i \in \VV}h_i(G)-G,
\end{align*}
where $G^*>0$ is the solution of
\begin{align}
\sum_{i \in \VV}\frac{\partial h_i(G^*)}{\partial G} = 1
\label{er01}
\end{align}
due to strict concaveness of $h_i$.
The socially optimal public good maximizes the aggregate payoff, and we define the aggregate payoff
$
\sum_{i \in \VV}h_i(G^*)-G^*
$
as the \textit{socially optimal payoff}. However, the socially optimal public good (as well as socially optimal payoff) is achieved only when $R \rightarrow \infty$ (Theorem 2 in~\cite{JM:00}). An infinite reward is apparently impractical. 
Moreover, existing works do not consider convex inequality constraint $g(s^*(R),R) \leq 0$ in the bi-level lottery, although it is essential in many engineering applications.
This paper aims to design a new incentive design to address the limitations.

\section{Optimal bi-level lottery design}\label{lin.sec:s_mechanism}
This section introduces a new practical scheme to achieve socially optimal payoff and satisfy convex inequality constraints.
In particular, a perturbed lottery model is introduced in Section~\ref{sec:ppm} and lower-level decision-making is presented in Section~\ref{sec:ppm1}. A new problem for the social planner is introduced in Section~\ref{sec:ppm2}. We highlight the differences from those in Section~\ref{lin.sec:Problem}.

\subsection{Perturbed payoff model}\label{sec:ppm}
Consider the \textit{perturbed payoff model} for player~$i$:
\begin{align}
U_i(s,R,c) \triangleq
\left\lbrace
\begin{array}{cc}
\frac{s_i-c_i}{\bar{s}-\bar{c}}R + h_i(\bar{s}-R) - s_i, &{\rm for } \ \bar{s} \geq R\\
0, & {\rm otherwise} \\
\end{array}
\right.
\label{e14}
\end{align}
where $c \triangleq  \{c_i\}_{i \in \VV}$ and $\bar{c} \triangleq \sum_{i \in \VV}c_i$, and $c_i$ is \textit{perturbation parameter}.
In~\eqref{e14}, $(R,c)$ is chosen by the social planner from a set ${\RR} \times {\CC}$ where ${\CC} \triangleq {\CC}_1 \times \cdots \times {\CC}_N$, and ${\CC}_i \triangleq  [0,\infty)$.
In the perturbed lottery, the social planner is able to choose reward $R$ and, at the same time, change the odds of winning by perturbing the individual investments.
In particular, perturbation parameter $c_i$ introduces an offset to the odds of winning but the aggregate portion remains one; i.e., $\sum_{i \in \VV}\frac{s_i-c_i}{\bar{s}-\bar{c}}=1$.
The perturbation is announced along with reward $R$ before holding a lottery.
The perturbed payoff model has been studied in~\cite{Bull:87,Gibbons:87}, where payoffs are functions of perturbed investment $s_i - c_i$ where $s_i$ is an effort made by player $i$, and $c_i$ is an external perturbation.
Likewise, perturbation $c$ is externally given by the social planner in the current paper, and it perturbs the portions of the reward to be received by players.
The external perturbation can be interpreted as an intervention of the social planner to achieve social optimum, and satisfy convex constraint $g(s^*(R,c),R) \leq 0$.

\subsection{Low-level decision making - Nash equilibrium}\label{sec:ppm1}
Given $R$, $c$, and $s_{-i}$, player $i$ chooses $s_i$ to maximize its own payoff:
$
\max_{s_i \in {\Ss}_i} U_i(s,R,c),
$ where $\Ss_i \triangleq  [0, \infty)$. Nash equilibrium $s^*(R,c)$ is dependent on $R$ and $c$. If the reward term $\frac{s_i^*(R,c)-c_i}{\bar{s}^*(R,c)-\bar{c}}R$ is negative, player $i$ is assumed to pay a fine $\frac{s_i^*(R,c)-c_i}{\bar{s}^*(R,c)-\bar{c}}R$ to the social planner.

\subsection{High-level decision making - Social optimum}\label{sec:ppm2}
As problem~\eqref{eq009}, the social planner wants to maximize the aggregate of perturbed payoffs as follows:
\begin{align}
&\begin{array}{c}
\underset{(R,c) \in {\RR} \times {\CC}}{\max}
\underset{i \in \VV}{\sum}
U_i(s^*(R,c),R,c)\\
{\rm s.t.}\quad g(s^*(R,c),R) \leq 0
\end{array}
\nnum\\
&=
\begin{array}{c}
\underset{(R,c) \in {\RR} \times {\CC}}{\max}
\underset{i \in \VV}{\sum}
h_i(G(R,c))-G(R,c)\\
{\rm s.t.}\quad g(s^*(R,c),R) \leq 0
\end{array}
\label{e172}
\end{align}
where $g: {\Ss} \times {\RR}\rightarrow {\real}^{m}$ remains identical to that in~\eqref{eq009} and the dependency of public good $G(R,c)\triangleq\bar{s}^*(R,c)-R$ on $R$ and $c$ is emphasized.
Since the objective function of~\eqref{e172} is identical to that of~\eqref{eq009},
it is maximized when $G(R,c)=G^*$ as well as constraint $g(s^*(R,c),R) \leq 0$ is absent.
There could be multiple optimal solutions for problem~\eqref{e172}.
%So the social planner aims to choose the optimal solution induced by minimal reward and perturbation.
So the social planner aims to choose the minimal reward and perturbation among the optimal solutions that satisfy the constraints $g(s^*(R,c),R) \leq 0$.
Considering this case, we formulate the following bi-level optimization problem:
\begin{align}
&\min_{(R,c) \in {\RR} \times {\CC}} R + \alpha\bar{c}\nnum\\
&\ {\rm s.t.}\begin{array}{l}
\ \ (R,c) \in {\argmax}
\underset{i \in \VV}{\sum}
U_i(s^*(R,c),R,c)\\
\quad {\rm s.t.}\quad g(s^*(R,c),R) \leq 0
\end{array}
\label{e17}
\end{align}
where constant $\alpha \geq 0$ represents a relative weight on the perturbation $\bar{c}$.

\section{Analysis of low-level Nash equilibrium}\label{lin.sec:an}
We study the properties of Nash equilibrium of problem~\eqref{e172} to solve problem~\eqref{e17}.
Theorem~\ref{the0} summarizes important properties of the perturbed lottery at Nash equilibrium. In particular, (P1) shows the existence and uniqueness of Nash equilibrium and (P2) demonstrates that any $(R,c)$ in the feasible set $\FF_{\eqref{e172}}$ of problem~\eqref{e172} satisfies $\bar{s}^*(R,c)-\bar{c}=G(R,c)+R-\bar{c}>0$ if there exists $(R,c)$ such that $G(R,c)=G^*$. (P2) helps to reduce the feasible set of interest.

\begin{theorem}
Suppose Assumption~\ref{nasm1} holds. Consider any pair $(R,c) \in {\RR} \times {\CC}$.
Then, the following properties hold at Nash equilibrium.
\begin{description}
\item[(P1)] Given any $c_i \geq 0$ for $\forall i \in \VV$ and $R>0$, there is a unique Nash equilibrium $s^*(R,c)$;
\item[(P2)] Given $|\VV| \neq 1$, any $(R,c)$ such that $G(R,c)=G^*$ satisfies $\bar{c}\leq G(R,c)+R$.
%Given $|\VV| \neq 1$, any $(R,c) \in \FF_{\eqref{e172}}$ satisfies $\bar{c}\leq G(R,c)+R$ if there exist a pair $(R,c) \in \FF_{\eqref{e172}}$ such that $G(R,c)=G^*$ holds.
\end{description}
\label{the0}
\end{theorem}
\begin{IEEEproof}
In the proof, we will drop the dependency of $G$, $s^*$, $U_i$, and $\VV_a$ on $R$ and $c$.

We first introduce the first order condition which must be satisfied at a Nash equilibrium:
\begin{align}
\frac{\partial U_i(s^*,R,c)}{\partial s_i} &= R \frac{\bar{s}^*-\bar{c}-(s_i^*-c_i)}{(\bar{s}^*-\bar{c})^2}+ \frac{\partial h_i(\bar{s}^*-R)}{\partial G} \nnum\\
&- 1 \leq 0
\label{lin.e:first}
\end{align}
for $\forall i \in \VV$.
If player $i$ is active; i.e., $s_i^*>0$, then equality holds.
If player $i$ is inactive; i.e., $s_i = 0$, strict inequality holds.
To prove the first order condition by contradiction, assume that $\frac{\partial U_i(s^*,R,c)}{\partial s_i}=\epsilon>0$. Then, by the Taylor series expansion, there exists a constant $\delta>0$ such that
\begin{align*}
U_i(s_i^*+\epsilon\delta,s^*_{-i},R,c)>U_i(s_i^*,s^*_{-i},R,c)+\epsilon\delta.
\end{align*}
This leads to a contradiction to the definition of Nash equilibrium.
The remaining part can be proven similarly.

(P1) Since $h_i$ is strictly increasing and strictly concave, there is $\xi_L > 0$ such that $\frac{\partial h_i(\xi)}{\partial G} < 1$ for all $\xi \geq \xi_L$. Consider any $R$, $c$ and $s_{-i}$. If $s_i$ is sufficiently large, then $U_i(s) < 0$. So there is $B_i(R,c) > 0$ such that $s_i^*(R,c) < B_i(R,c)$. Hence, $s^*(R,c)$ is identical to the maximizer of the game: $\max_{s_i}U_i(s)$ s.t. $s_i\in[0,B_i(R,c)]$. In this problem, the payoff functions are concave and the decision variables lie in compact sets. Hence, $s^*(R,c)$ exists.
The uniqueness of Nash equilibrium can be proven by similar arguments of Lemma 3 in~\cite{JM:00}.

(P2) We show by contradiction. Assume $\bar{s}^*-\bar{c}=G+R-\bar{c}<0$ and $G=G^*$.
The aggregate of the first order conditions of active players is
\begin{align*}
\sum_{i \in \VV_a}\frac{\partial U_i(s^*)}{\partial s_i} &= \frac{R(|\VV_a|-1)}{R+G-\bar{c}} +\frac{\sum_{i \in \VV \setminus \VV_a}(s_i^*-c_i)}{(R+G-\bar{c})^2}R\nnum\\
&+ \sum_{i \in \VV_a}\frac{\partial h_i(G)}{\partial G} - |\VV_a| = 0.
\end{align*}
Since $\frac{\partial h_i(G)}{\partial G}>0$ and $s_i=0$ for $i \in \VV \setminus \VV_a$, we have
\begin{align}
&\frac{R(|\VV_a|-1)}{R+G-\bar{c}} -\frac{\sum_{i \in \VV \setminus \VV_a}c_i}{(R+G-\bar{c})^2}R+ \sum_{i \in \VV}\frac{\partial h_i(G^*)}{\partial G} - |\VV_a|\nnum\\
&=-\frac{(G-\bar{c})(|\VV_a|-1)}{R+G-\bar{c}}-\frac{\sum_{i \in \VV \setminus \VV_a}c_i}{(R+G-\bar{c})^2}R \geq 0
\label{3rd.001}
\end{align}
which never holds when $|\VV_a| \neq 1$ because $G-\bar{c}<R+G-\bar{c}<0$ and $c_i \geq 0$.

If $|\VV_a|=1$, then~\eqref{3rd.001} holds when $c_i=0$ for $\forall i \in \VV \setminus \VV_a$.
The first order condition of $k \in \VV_a$ is
\begin{align*}
\frac{\partial U_k(s^*)}{\partial s_k} &= R \frac{\bar{s}^*-\bar{c}-(s_k^*-c_k)}{(\bar{s}^*-\bar{c})^2}+ \frac{\partial h_k(\bar{s}^*-R)}{\partial G} - 1\nnum\\
&=\frac{\partial h_k(\bar{s}^*-R)}{\partial G} - 1=0
\end{align*}
where $c_k = \bar{c}$ is applied. Thus, $\frac{\partial h_k(G^*)}{\partial G} =1$ and
$\sum_{i \in \VV}\frac{\partial h_i(G^*)}{\partial G}$ $>1$.
This contradicts the definition of $G^*$.
\end{IEEEproof}

%The proof of Theorem~\ref{the0} is motivated by the technique developed in~\cite{healy2012designing}.
If the convex constraint $g(s^*(R,c),R)\leq0$ is absent in problem~\eqref{e172}, there always exists $(R,c)$ such that $G(R,c)=G^*$ and thus (P2) holds (which will be shown later). In other words, (P2) holds if the optimal value of problem~\eqref{e172} remains $\sum_{i \in \VV}h_i(G^*)-G^*$ with/without the convex constraint $g(s^*(R,c),R)\leq0$.

According to (P2), we only focus on the feasible set with the constraint $\bar{c}\leq G(R,c)+R$ in problem~\eqref{e172}.
This constraint makes us derive the following properties further while it does not restrict the choice of $(R,c)$.
Theorem~\ref{the1} summarizes the derived properties, and these properties are essential to solve bi-level optimization problem~\eqref{e17}. Furthermore, the properties reduce to those of unperturbed lottery in Section~\ref{lin.sec:Problem} when $c_i=0$.
(P3) indicates that public good $G(R,c)$ is bounded by $\bar{c}$ and $G^*$, and it is increasing in $(R,c)$ when all the players are active.
(P4) shows that all the players are active if reward $R$ is greater than a certain threshold, and there exists a lower bound of $s^*_{i}(R,c)$, which is a strictly increasing function in $R$.
Moreover, in some cases, $s_i^*(R,c)$ is strictly increasing in $(R,c)$.
(P5) quantifies the price of anarchy~\cite{Koutsoupias:99} which is the ratio between the socially optimal payoff and the aggregate payoff induced by the corresponding Nash equilibrium. The lower and upper bounds of the price of anarchy reveal possible efficiency losses due to selfishness of players, and they can be quantified without explicitly calculating Nash equilibrium.

The following notations are used in Theorem~\ref{the1}. Function $sgn(\cdot)$ is a sign function.
The value $R_L(c)$ is the unique solution of $\frac{R_L(c)}{R_L(c)+G^U-\bar{c}} = \max_{i \in \VV} \{1 - \frac{\partial h_i(G^U)}{\partial G}\}$, where $G^U \triangleq \max\{G^*,\bar{c}\}$.
Define player $i$ who invests non-zero wealth $s_i^*(R,c) >0$ as an active player and define $\VV_a(R,c) \triangleq \{i \in \VV | s_i^*(R,c)>0\}$ as the set of all the active players.
Lastly, if $\bar{c}\leq G^*$, then
\begin{align*}
\underline{G}(R,c) &\triangleq H^{-1}(\frac{(|\bar{\VV}_a(R,c)|-1)(G^L-\bar{c})}{R+G^U-\bar{c}} +1)\nnum\\
\bar{G}(R,c) &\triangleq H^{-1}(\frac{(N-1)(G^U-\bar{c})}{R+G^L-\bar{c}} +1)
\end{align*}
otherwise
\begin{align*}
\bar{G}(R,c) &\triangleq H^{-1}(\frac{(|\bar{\VV}_a(R,c)|-1)(G^L-\bar{c})}{R+G^L-\bar{c}} +1)\nnum\\
\underline{G}(R,c) &\triangleq H^{-1}(\frac{(N-1)(G^U-\bar{c})}{R+G^U-\bar{c}} +1)
\end{align*}
where $G^L \triangleq \min\{G^*,\bar{c}\}$, $H(G) \triangleq \sum_{i \in \VV}\frac{\partial h_i(G)}{d G}$ and $\bar{\VV}_a(R,c)$ is the number of players who satisfy
$\frac{R}{R+G^U-\bar{c}}+\frac{\partial h_i(G^U)}{\partial G}-1> 0$.
%$\frac{R}{R+G^U-\bar{c}}+\frac{\partial h_i(G^U)}{\partial G}-1> 0$.
Note that $H: {\real}_{\geq 0} \rightarrow Y$ is invertible on codomain $Y \triangleq (0,H(0)]$ because $H$ is a strictly decreasing and continuous.

\begin{theorem}
Suppose Assumption~\ref{nasm1} holds. Consider any pair $(R,c) \in {\RR} \times {\CC}$ such that $\bar{c}\leq G(R,c)+R$. Then, the following properties hold at Nash equilibrium.
\begin{description}
\item[(P3)] It holds that $\bar{c} \leq G(R,c) \leq G^*$ or $G^*\leq G(R,c) \leq \bar{c}$. If $|\VV_a(R,c)|=N$, then $sgn(G^*-\bar{c})\frac{d G(R,c)}{d R}\geq0$, and $\frac{d G(R,c)}{d c_i}>0$ where equality holds if and only if $\bar{c}=G^*$;
\item[(P4)] If $R > R_L(c)$, then $s_i^*(R,c) \geq c_i + R \big(\frac{R}{R+G^U-\bar{c}}+\frac{\partial h_i(G^U)}{\partial G}-1 \big) > 0$ where the lower bound is strictly increasing in $R$ without bound.
If $|\VV_a(R,c)|=N$, there is some $i \in \VV$ such that $\frac{d s_i^*(R,c)}{d R}>0$.
%Moreover, if $|\VV_a(R,c)|=N$, $h_i=h_j$ for $\forall i,j \in \VV$, then $\frac{d s_i^*(R,c)}{d R}\geq \frac{1}{N}$, and $\frac{d s_i^*(R,c)}{d c_i}>0$ for $\forall i$;
\item[(P5)] Price of anarchy
${\rm PoA}(R,c) \triangleq \frac{\max_{s \in {\Ss}}\sum_{i \in \VV}U_i(s)}{\sum_{i \in \VV}U_i(s^*(R,c),R,c)}$ is characterized by
\begin{align*}
&\frac{\sum_{i \in \VV}h_i(G^*)-G^*}{\sum_{i \in \VV}h_i(\underline{G}(R,c))-\underline{G}(R,c)}\leq
{\rm PoA}(R,c)\nnum\\
&\leq\frac{\sum_{i \in \VV}h_i(G^*)-G^*}{\sum_{i \in \VV}h_i(\bar{G}(R,c))-\bar{G}(R,c)}.
\end{align*}
If $\bar{c}=0$, it holds that ${\rm PoA}>1$ for any $R<\infty$ and $\lim_{R \rightarrow \infty}{\rm PoA}(R,0)=1$.
\end{description}
\label{the1}
\end{theorem}
\begin{IEEEproof}
In the proof, we will drop the dependency of $G$, $\underline{G}$, $\bar{G}$, $s^*$, $U_i$, $\VV_a$, $R_L$ and ${\rm PoA}$ on $R$ and $c$.

(P3) Assume $G\leq \bar{c}$. The aggregate of the first order conditions~\eqref{lin.e:first} is
\begin{align}
\sum_{i \in \VV}\frac{\partial U_i(s^*)}{\partial s_i} &= \frac{R(N-1)}{R+G-\bar{c}} + \sum_{i \in \VV}\frac{\partial h_i(G)}{\partial G} - N \leq 0
\label{2nd.ee:01}
\end{align}
and thus we have
$
\sum_{i \in \VV}\frac{\partial h_i(G)}{\partial G} \leq 1 =\sum_{i \in \VV}\frac{\partial h_i(G^*)}{\partial G}.
$
This implies $G^*\leq G \leq\bar{c}$ due to strict concaveness of $h_i$.

Now assume $G\geq \bar{c}$. The aggregate of the first order conditions~\eqref{lin.e:first} of active players:
\begin{align*}
\sum_{i \in \VV_a}\frac{\partial U_i(s^*)}{\partial s_i} &= \frac{R(|\VV_a|-1)}{R+G-\bar{c}} +\frac{\sum_{i \in \VV \setminus \VV_a}(s_i^*-c_i)}{(R+G-\bar{c})^2}R\nnum\\
&+ \sum_{i \in \VV_a}\frac{\partial h_i(G)}{\partial G} - |\VV_a| = 0.
\end{align*}
Note that $s_i^*=0$ for $i \in \VV \setminus \VV_a$. By the fact that $h_i$ is a strictly increasing function, it becomes
\begin{align}
&\sum_{i \in \VV}\frac{\partial h_i(G)}{\partial G}  \geq -\frac{R(|\VV_a|-1)}{R+G-\bar{c}} + |\VV_a| +\frac{\sum_{i \in \VV \setminus \VV_a}c_i}{(R+G-\bar{c})^2}R\nnum\\
& = \frac{(|\VV_a|-1)(G-\bar{c})}{R+G-\bar{c}} +\frac{\sum_{i \in \VV \setminus \VV_a}c_i}{(R+G-\bar{c})^2}R+1\nnum\\
& = \frac{(|\VV_a|-1)(G-\bar{c})}{R+G-\bar{c}} +\frac{\sum_{i \in \VV \setminus \VV_a}c_i}{(R+G-\bar{c})^2}R+ \sum_{i \in \VV}\frac{\partial h_i(G^*)}{\partial G}.
\label{2nd.ee:03}
\end{align}
Because $\frac{(|\VV_a|-1)(G-\bar{c})}{R+G-\bar{c}}\geq 0$ and $c_i \geq 0$,~\eqref{2nd.ee:03} implies $G \leq G^*$ by strict concaveness of $h_i$. Thus, $\bar{c} \leq G \leq G^*$.

Now we consider the case with $|\VV_a|=N$.
Since all the players are active the aggregate first order condition~\eqref{2nd.ee:01} holds with equality where $\sum_{i \in \VV}\frac{\partial U_i(s^*)}{\partial s_i}$ can be regarded as an implicit function of $(s^*,R,c)$.
We apply the implicit function theorem (Theorem 1.3.1 in~\cite{krantz2012implicit}) to~\eqref{2nd.ee:01}
\begin{align*}
-\frac{\partial (\sum_{i \in \VV}\frac{\partial U_i(s^*)}{\partial s_i})}{\partial G}\frac{d G}{d R}=
\frac{\partial (\sum_{i \in \VV}\frac{\partial U_i(s^*)}{\partial s_i})}{\partial R}
\end{align*}
and obtain
\begin{align}
\frac{d G}{d R}=-\frac{(G-\bar{c})(N-1)}{(R+G-\bar{c})^2\sum_{i \in \VV}\frac{\partial^2 h_i(G)}{\partial G^2}-R(N-1)}.
\label{ee004}
\end{align}
Thus, $\frac{d G}{d R}\geq0$ if $\bar{c} \leq G \leq G^*$ and $\frac{d G}{d R}\leq0$ if $G^* \leq G \leq \bar{c}$.
It holds that $\frac{d G}{d R}=0$ if and only if $G=\bar{c}$.

We will show that $G=\bar{c}$ if and only if $\bar{c}=G^*$. If $\bar{c}=G^*$, then $G=\bar{c}$ because $\bar{c} \leq G \leq G^*$. We now prove that if $G=\bar{c}$ then $\bar{c}=G^*$. Assume $G=\bar{c}$, then aggregate first order condition~\eqref{2nd.ee:01} yields
\begin{align*}
\sum_{i \in \VV}\frac{\partial U_i(s^*)}{\partial s_i} &= \sum_{i \in \VV}\frac{\partial h_i(\bar{c})}{\partial G} -1 = 0.
\end{align*}
The unique solution is $\bar{c}=G^*$.

We proceed to prove $\frac{d G}{d c_i}>0$.
By applying the implicit function theorem to~\eqref{2nd.ee:01}, we have
\begin{align*}
-\frac{\partial (\sum_{i \in \VV}\frac{\partial U_i(s^*)}{\partial s_i})}{\partial G}\frac{d G}{d c_i}=
\frac{\partial (\sum_{i \in \VV}\frac{\partial U_i(s^*)}{\partial s_i})}{\partial c_i}
\end{align*}
and obtain
\begin{align}
\frac{d G}{d c_i}=-\frac{R(N-1)}{(R+G-\bar{c})^2\sum_{i \in \VV}\frac{\partial^2 h_i(G)}{\partial G^2}-R(N-1)}>0.
\label{ee005}
\end{align}

(P4) By $G \leq \max\{G^*,\bar{c}\} \triangleq G^U$ and concaveness of $h_i$, first order condition~\eqref{lin.e:first} yields
\begin{align}
&\frac{\partial U_i(s^*)}{\partial s_i} = R \frac{\bar{s}^*-\bar{c}-(s_i^*-c_i)}{(\bar{s}^*-\bar{c})^2} + \frac{\partial h_i(\bar{s}^*-R)}{\partial G} - 1 \nnum\\
& \quad \quad \quad \geq \frac{R}{\bar{s}^*-\bar{c}} - R\frac{s_i^*-c_i}{(\bar{s}^*-\bar{c})^2} + \frac{\partial h_i(G^U)}{\partial G}-1.
\label{2nd.e:58}
\end{align}
Assume $s_i^* < c_i$, then with $R> R_L$,
\begin{align*}
\frac{\partial U_i(s^*)}{\partial s_i} &> \frac{R_L}{R_L+G^U-\bar{c}} + \frac{\partial h_i(G^U)}{\partial G}-1=0.
\end{align*}
This contradicts the first order condition, and thus $s_i^* \geq c_i$. With $G^U \geq \bar{c}$,~\eqref{2nd.e:58} becomes
\begin{align*}
\frac{\partial U_i(s^*)}{\partial s_i} &\geq \frac{R}{R+G^U-\bar{c}} - \frac{s_i^*-c_i}{R} + \frac{\partial h_i(G^U)}{\partial G}-1.
\end{align*}
If $s_i^* < c_i + R \big(\frac{R}{R+G^U-\bar{c}}+\frac{\partial h_i(G^U)}{\partial G}-1\big)$, then $\frac{\partial U_i(s^*)}{\partial s_i}>0$, a contradiction to the first order condition.
Therefore $s_i^* \geq c_i + R \big(\frac{R}{R+G^U-\bar{c}}+\frac{\partial h_i(G^U)}{\partial G}-1 \big)$ and the lower bound is strictly positive, because $R > R_L$.

We now proceed to prove that the bound $L_i(R,c) \triangleq c_i + R \big(\frac{R}{R+G^U-\bar{c}}+\frac{\partial h_i(G^U)}{\partial G}-1 \big)$ is a strictly increasing in $R$ without bound.
By taking derivative of the bound, we have
\begin{align*}
\frac{\partial L_i}{\partial R} &= \frac{R}{R+G^U-\bar{c}}+\frac{\partial h_i(G^U)}{\partial G}-1+ R \frac{G^U-\bar{c}}{(R+G^U-\bar{c})^2}
\end{align*}
which is strictly greater than $0$ since $G^U \geq \bar{c}$ and $R> R_L$.
Moreover, function $L_i$ keeps increasing without bound as $R$ increases because $\lim_{R \rightarrow \infty} \frac{\partial L_i}{\partial R} = \frac{\partial h_i(G^U)}{\partial G}>0$.

Now we will consider the case with $|\VV_a|=N$.
We will show that there is at least one $i$ such that $\frac{d s_i^*}{d R}>0$ holds.
Since all the players are active, the first order condition~\eqref{lin.e:first} holds with equality
$\frac{\partial U_i(s^*)}{\partial s_i} =0$ where $\frac{\partial U_i(s^*)}{\partial s_i}$ can be regarded as an implicit function of $(s^*,R,c)$. By the implicit function theorem, relation
\begin{align*}
&-
\left[
\begin{array}{ccc}
\frac{\partial^2 U_1 (s^*)}{\partial s_1^2}&\cdots&\frac{\partial^2 U_1 (s^*)}{\partial s_1 \partial s_N}\\
\vdots&\ddots&\vdots\\
\frac{\partial^2 U_N (s^*)}{\partial s_N \partial s_1}&\cdots&\frac{\partial^2 U_N (s^*)}{\partial s_N^2}
\end{array}
\right]
\left[
\begin{array}{c}
\frac{d s_1^*}{d R}\\
\vdots\\
\frac{d s_N^*}{d R}\\
\end{array}
\right]=\left[
\begin{array}{ccc}
\frac{\partial^2 U_1 (s^*)}{\partial s_1 \partial R}\\
\vdots\\
\frac{\partial^2 U_N (s^*)}{\partial s_N \partial R}\\
\end{array}
\right]
\end{align*}
holds where
\begin{align*}
\frac{\partial^2 U_i(s^*)}{\partial s_i^2} &= -2R \frac{\bar{s}^*-\bar{c}-(s_i^*-c_i)}{(\bar{s}^*-\bar{c})^3} + \frac{\partial^2 h_i(\bar{s}^*-R)}{\partial G^2}<0\nnum\\
\frac{\partial^2 U_i(s^*)}{\partial s_i \partial s_j} &= -R \frac{\bar{s}^*-\bar{c}-2(s_i^*-c_i)}{(\bar{s}^*-\bar{c})^3} + \frac{\partial^2 h_i(\bar{s}^*-R)}{\partial G^2}\nnum\\
\frac{\partial^2 U_i(s^*)}{\partial s_i \partial R} &= \frac{\bar{s}^*-\bar{c}-(s_i^*-c_i)}{(\bar{s}^*-\bar{c})^2} - \frac{\partial^2 h_i(\bar{s}^*-R)}{\partial G^2}>0.
\end{align*}
If we choose $k={\rm argmin}_{i \in \VV} \ s_i^*$, it holds that $\frac{\partial^2 U_k(s^*)}{\partial s_k \partial s_j}\leq0$ because $s_k^*>c_k$.
Therefore, the relation
\begin{align*}
-\sum_{j \in \VV} \frac{\partial^2 U_k (s^*)}{\partial s_k \partial s_j}\frac{d s_j^*}{d R} = \frac{\partial^2 U_k (s^*)}{\partial s_k \partial R}>0
\end{align*}
implies that there is at least one $j$ such that $\frac{d s_j^*}{d R}>0$.

(P5) It holds that
\begin{align}
&\frac{(|\VV_a|-1)(G-\bar{c})}{R+G-\bar{c}}
+\frac{R\sum_{i \in \VV \setminus \VV_a}c_i}{(R+G-\bar{c})^2}+1 \nnum\\
&\leq \sum_{i \in \VV}\frac{\partial h_i(G)}{\partial G}\leq \frac{(N-1)(G-\bar{c})}{R+G-\bar{c}} +1
\label{ew105}
\end{align}
where the lower bound can be found from~\eqref{2nd.ee:03} and the upper bound can be obtained from~\eqref{2nd.ee:01}:
\begin{align*}
\sum_{i \in \VV}\frac{\partial h_i(G)}{\partial G} &\leq -\frac{R(N-1)}{R+G-\bar{c}} + N = \frac{(N-1)(G-\bar{c})}{R+G-\bar{c}}+ 1.
\end{align*}
If $i \in \bar{\VV}_a$, then $i \in \VV_a$ because it holds that $s_i^* \geq c_i + \frac{(G+R-\bar{c})^2}{R} \big(\frac{R}{R+G-\bar{c}}+\frac{\partial h_i(G)}{\partial G}-1\big)
\geq c_i + R \big(\frac{R}{R+G^U-\bar{c}}+\frac{\partial h_i(G^U)}{\partial G}-1 \big)$ by inequality~\eqref{2nd.e:58}.
Given $\bar{c}\leq G^*$, inequality~\eqref{ew105} implies that
$
\bar{G} \leq G \leq \underline{G}
$
because $H$ is a strictly decreasing function.
It holds that $\underline{G}\leq G^*$ because $G^*=H^{-1}(1)$ and $H^{-1}$ is also strictly decreasing.
Since $\sum_{i \in \VV}h_i(G)-G$ is strictly increasing in $G \in [0,G^*]$ and has a maximum at $G=G^*$, we have
\begin{align}
\sum_{i \in \VV}h_i(\bar{G})-\bar{G} \leq \sum_{i \in \VV}U_i(s^*) \leq \sum_{i \in \VV}h_i(\underline{G})-\underline{G}.
\label{ew106}
\end{align}
Dividing
\begin{align*}
\max_{s \in {\Ss}}\sum_{i \in \VV}U_i(s) &= \max_{s \in {\Ss}}\sum_{i \in \VV}h_i(\bar{s}-R)-(\bar{s}-R) \nnum\\
&= \sum_{i \in \VV}h_i(G^*)-G^*
\end{align*}
by~\eqref{ew106} yields the desired result. Now we proceed to prove that ${\rm PoA}>1$ with any $R<\infty$ if $c=0$, but it holds that $\lim_{R \rightarrow \infty}{\rm PoA}(R,0)=1$. It can be shown that $\underline{G} = H^{-1}(\frac{(|\bar{\VV}_a|-1)\bar{G}}{R+G^U-\bar{c}} +1)<H^{-1}(1)=G^*$ where $\bar{G}\neq0$. Therefore, $1<\frac{\sum_{i \in \VV}h_i(G^*)-G^*}{\sum_{i \in \VV}h_i(\underline{G})-\underline{G}}\leq {\rm PoA}$ with any $R<\infty$. Moreover, as $R \rightarrow \infty$, it holds that $\lim_{R \rightarrow \infty} \bar{G} = \lim_{R \rightarrow \infty} \underline{G} = H^{-1}(1)=G^*$. Therefore, we can conclude that, if $\bar{c}\leq G^*$, then $\lim_{R \rightarrow \infty}{\rm PoA} =1$.
The properties with $\bar{c}> G^*$ can be proven similarly. We omit its details.
\end{IEEEproof}

A pair $(R,c) = (G^*,G^*)$ satisfies $G(R,c)+R-\bar{c} = G(R,c)\geq0$ and, by (P3), it holds that $G(R,c)=G^*$. Therefore, there always exists at least one pair $(R,c)$ such that $G(R,c)=G^*$ if the convex constraint $g(s^*(R,c),R)\leq0$ is absent.
(P3) shows that payoff~\eqref{e14} does not have discontinuity because $\bar{s}^*(R,c)>\bar{c}$.
Remind that ${\rm PoA}(R,c)=1$ if and only if $G(R,c)=G^*$. So (P5) indicates that it is impossible to achieve optimality $G(R,c)=G^*$ with a finite reward when perturbations are not allowed; i.e., there is no finite maximizer of problem~\eqref{eq009}. Price of anarchy is identical to Price of stability~\cite{anshelevich2008price} which represents the ratio between
the socially optimal payoff and the aggregate payoff induced by the best Nash equilibrium because there exists a unique Nash equilibrium by (P1). % (P4) shows that problem~\eqref{e172} is not well-defined neither because $s_i^*(R,c)$ increases unbounded as $R$ increases.

Some properties of Theorems~\ref{the0} and~\ref{the1} reduce to those in~\cite{JM:00} where perturbations are absent. In particular, (P1) reduces to Proposition 2 of~\cite{JM:00}, where an unperturbed lottery has a unique Nash equilibrium. (P5) is consistent with Theorem 2 in~\cite{JM:00}; i.e., given any $\epsilon>0$, there exists $R$ such that ${\rm PoA}(R,0) \leq 1+\epsilon$. The lower and upper bounds of price of anarchy are newly derived in this paper and they can be calculated without finding the Nash equilibrium. Additionally, (P3) and (P4) are new and reveal the properties of public goods and investment, respectively.

\section{Convex approximation of high-level social optimum}\label{leastP}
Problem~\eqref{e17} is a bi-level optimization problem. In general, this class of problems is computationally challenging. In particular, papers~\cite{bard1991some,jeroslow1985polynomial,vicente1994descent} show that bi-level linear programs are NP-hard.
Given the computational hardness, certain relaxations of problem~\eqref{e17} are needed in order to find computationally efficient solvers. We will leverage Theorem~\ref{the1} to show that the following problem is a convex reformulation for problem~\eqref{e17}. We will also show that, under certain mild conditions, the approximation gap is zero. Consider
\begin{align}
&\min_{(R,c) \in {\RR} \times {\CC}} R+\alpha G^*\nnum\\
&\ {\rm s.t.} \ \bar{c} =G^*,\nnum\\
& \quad \ g(c_1 +R\frac{\partial h_1(G^*)}{\partial G},\cdots,c_N +R\frac{\partial h_N(G^*)}{\partial G},R) \leq 0.
\label{new.e0}
\end{align}
The problem~\eqref{new.e0} is convex.
The objective function is affine, constraint $\bar{c}=G^*$ is also affine.
Constraint
\begin{align}
&g(c_1 +R\frac{\partial h_1(G^*)}{\partial G},\cdots,c_N +R\frac{\partial h_N(G^*)}{\partial G},R) \leq 0
\label{ew004}
\end{align}
is convex because a composition of convex function with affine functions preserves convexity
where $g$ is a convex function by Assumption~\ref{nasm3} and $c_i +R\frac{\partial h_i(G^*)}{\partial G}$ is an affine function.
Feasible set ${\RR} \times {\CC}$ is convex because ${\RR} = (0,\infty)$, ${\CC}_i = [0, G^*]$ are convex sets.
%and Cartesian products preserve convexity.
The following theorem shows that problem~\eqref{new.e0} is a convex reformulation of problem~\eqref{e17} if there exists a pair $(R,c)$ such that $G(R,c)=G^*$.
%This condition always holds if the convex constraint $g(s^*,R,c)\leq0$ does not prevent the social planner from achieving the social optimum $G(R,c)=G^*$.

Sets $\FF_{\eqref{e17}}$, and $\FF_{\eqref{new.e0}}$ denote the feasible sets of problems~\eqref{e17}, and~\eqref{new.e0} respectively.
Likewise, we define $p^*_{\eqref{e17}}$, and $p^*_{\eqref{new.e0}}$ as the optimal values of problems~\eqref{e17}, and~\eqref{new.e0} respectively. 
\begin{theorem}
Assume that there is $(R,c) \in \FF_{\eqref{e172}}$ such that $G(R,c)=G^*$.
Under Assumptions~\ref{nasm1}, and~\ref{nasm3}, it holds that $\FF_{\eqref{new.e0}} = \FF_{\eqref{e17}}$ and $p^*_{\eqref{new.e0}}= p^*_{\eqref{e17}}$.
\label{the51}
\end{theorem}
\begin{IEEEproof}
Notice that there is $(R,c) \in \FF_{\eqref{e172}}$ such that $G(R,c)=G^*$, and this implies that we can only focus on a feasible set with constraint $G+R \geq \bar{c}$ according to (P2); i.e., all the analysis in Section~\ref{lin.sec:an} is valid.
In the proof, we will drop the dependency of $G$, $s^*$, and $U_i$ on $R$ and $c$.
The proofs are divided into two claim statements.
\begin{claim}
$\FF_{\eqref{new.e0}}$ is a subset of $\FF_{\eqref{e17}}$.
\label{clm1}
\end{claim}
%\textsc{Claim 1.} \emph{$\FF_{\eqref{new.e0}}$ is a subset of $\FF_{\eqref{e17}}$.}\\
\begin{IEEEproof}
Assume that $\FF_{\eqref{new.e0}}$ is non-empty and we pick any $(R,c) \in \FF_{\eqref{new.e0}}$. We will show that such the pair $(R,c)$ satisfies all the constraints in~\eqref{e17}; i.e., $(R,c) \in \FF_{\eqref{e17}}$.

The constraint $\bar{c}=G^*$ implies that $G^* = \bar{c} = G$ by (P3). Therefore, it holds that $G=G^*$ which implies $(R,c) \in \argmax \sum_{i \in \VV} U_i(s^*(R,c),R,c)$ in~\eqref{e17}.

Using $\bar{s}^*-R=G^*=\bar{c}$, the first order condition yields
\begin{align}
\frac{\partial U_i(s^*)}{\partial s_i} &= R \frac{\bar{s}^*-\bar{c}-(s_i^*-c_i)}{(\bar{s}^*-\bar{c})^2} + \frac{\partial h_i(\bar{s}^*-R)}{\partial G} - 1 \nnum\\
&= -\frac{s_i^*-c_i}{R} + \frac{\partial h_i(G^*)}{\partial G}  \leq 0.
\label{eqn09}
\end{align}
This inequality implies that
\begin{align}
s_i^* \geq c_i +R\frac{\partial h_i(G^*)}{\partial G}>0
\label{eqn2}
\end{align}
because $G^*>0$ and $R>0$.
Since the players are active, equality holds in the first order condition~\eqref{eqn09} as well as~\eqref{eqn2}:
\begin{align}
s_i^* = c_i +R\frac{\partial h_i(G^*)}{\partial G}.
\label{ew006}
\end{align}
Therefore, constraint~\eqref{ew004} implies
\begin{align}
&g(s^*,R) \leq 0.
\label{ew005}
\end{align}
%Lastly, the constraint $\bar{c} \leq R+G$ holds obviously.
Therefore, $(R,c) \in {\FF}_{\eqref{e17}}$.
The statement holds because we pick arbitrary $(R,c) \in {\FF}_{\eqref{new.e0}}$.
\end{IEEEproof}

Claim~\ref{clm1} shows that $\FF_{\eqref{new.e0}} \subseteq \FF_{\eqref{e17}}$.
The objective function of~\eqref{new.e0} is $\min_{(R,c)} R+\alpha G^*=\min_{(R,c)} R+\alpha \bar{c}$.
Therefore, solution $p^*_{\eqref{new.e0}}$ is an overestimate of $p^*_{\eqref{e17}}$.
We now proceed to prove that $\FF_{\eqref{e17}} \subseteq \FF_{\eqref{new.e0}}$ and thus $p^*_{\eqref{e17}}=p^*_{\eqref{new.e0}}$.
\begin{claim}
$\FF_{\eqref{e17}}$ is a subset of $\FF_{\eqref{new.e0}}$.
\label{clm2}
\end{claim}
\begin{IEEEproof}
Assume that $\FF_{\eqref{e17}}$ is non-empty and we pick any $(R,c) \in \FF_{\eqref{e17}}$.
We will show that the pair satisfies all the constraints in~\eqref{new.e0}.

Recall that we assume that there exist $(R,c)$ such that $G=G^*$, and all the other constraints are satisfied. 
We now prove $\bar{c}=G^*$ by contradiction. Assume that there exist pair $(R,c)$ such that $G=G^*$ but $\bar{c} \neq G^*$.
By~\eqref{ew105},
\begin{align*}
&\frac{(|\VV_a|-1)(G^*-\bar{c})}{R+G^*-\bar{c}} +1
+\frac{R\sum_{i \in \VV \setminus \VV_a}c_i}{(R+G^*-\bar{c})^2}\leq \sum_{i \in \VV}\frac{\partial h_i(G^*)}{\partial G},
\end{align*}
which holds only if $G^* < \bar{c}\leq R+G^*$ or $|\VV_a|=1$.
Let us consider the first case.
From~\eqref{2nd.ee:01}, we have
\begin{align*}
\sum_{i \in \VV}\frac{\partial U_i(s^*)}{\partial s_i} &= \frac{R(N-1)}{R+G^*-\bar{c}} + \sum_{i \in \VV}\frac{\partial h_i(G^*)}{\partial G} - N \leq 0
\end{align*}
which holds only if $\bar{c}\leq G^*$, a contradiction. Therefore, it must hold that $|\VV_a|=1$.
First order condition~\eqref{lin.e:first} for $i \in \VV_a$ must hold with equality. However, we have
%\begin{align*} \frac{\partial U_i(s^*)}{\partial s_i}=-R\frac{\bar{c}-c_i}{(G^*+R-\bar{c})^2}+\frac{\partial h_i(G^*)}{\partial G}-1<0 \end{align*}
\begin{align*}
\frac{\partial U_i(s^*)}{\partial s_i}&=-R\frac{\bar{c}-c_i}{(G^*+R-\bar{c})^2}+\frac{\partial h_i(G^*)}{\partial G}-1\nnum\\
&\leq\sum_{j \in \VV \setminus i}\frac{\partial h_j(G^*)}{\partial G}<0
\end{align*}
which contradicts to the first order condition, where we applied $\sum_{i \in \VV}\frac{\partial h_i(G^*)}{\partial G}=1$.
Therefore, $\bar{c}=G^*$. Note that if $\bar{c}=G^*$, then $G=G^*$ by (P3).

First order condition with $\bar{c}=G^*$
\begin{align*}
\frac{\partial U_i(s^*)}{\partial s_i}=-\frac{s_i^*}{R}+\frac{c_i}{R}+\frac{\partial h_i(G^*)}{\partial G}\leq0
\end{align*}
implies $s_i^*>0$, which holds for $\forall i \in \VV$; i.e., $|\VV_a|=N$.

Using the first order condition~\eqref{eqn2}, we can derive
\begin{align}
s_i^* = c_i +R\frac{\partial h_i(G^*)}{\partial G}
\label{eqn3}
\end{align}
where equality holds because all the players are active.
By plugging~\eqref{eqn3} into constraint~\eqref{ew005}, we obtain~\eqref{ew004}.
%Since $\bar{c}=G=G^*$, the first constraint $\bar{c}=G^*$ in~\eqref{new.e0} holds.
% Constraint $R \geq G^*$ is satisfied since $R \geq \bar{c}=G^*$.
Therefore, $(R,c) \in {\FF}_{\eqref{new.e0}}$.
The statement holds because we pick arbitrary $(R,c) \in {\FF}_{\eqref{e17}}$.
\end{IEEEproof}
Claim~\ref{clm1} and Claim~\ref{clm2} imply that $\FF_{\eqref{e17}} = \FF_{\eqref{new.e0}}$. The objective functions are equivalent to each other because $\bar{c} = G^*$ for the both feasible sets. Thus it holds that $p^*_{\eqref{e17}}=p^*_{\eqref{new.e0}}$.
\end{IEEEproof}

In Theorem~\ref{the51}, non-convex optimization problem~\eqref{e17} is approximated by (or equivalent to) a convex optimization problem~\eqref{new.e0}, which can be efficiently solved~\cite{boyd2004convex}.
In particular, with $\bar{c}=G^*$, we could obtain a constant public good $G(R,c)=G^*$ by (P3) in Theorem~\ref{the1}, which sequentially results in the replacement of potentially non-concave function $s_i^*(R,c)$ with a linear function~\eqref{ew006}.
This condition is a sufficient and necessary condition for the social optimum.

\begin{remark}
The optimal bi-level lottery design may not guarantee individual rationality; i.e., $U_i(s^*,R,c) \geq 0$. However, as long as there exists $(R,c)$ such that $G(R,c)=G^*$, individual rationality can be readily ensured by adding the convex constraints $g$ as follows:
\begin{align*}
&g_i(s^*(R,c),R)=-(s_i^*-c_i+Rh_i(G^*)-Rs_i^*)\leq0
\end{align*}
for $\forall i \in \VV$
where $h_i(G^*)$ is a constant.
By applying $G^*=\bar{c}$ into $U_i(s^*,R,c) \geq 0$ in~\eqref{e14}, one can show that the above condition is equivalent to $U_i(s^*,R,c) \geq 0$ for $\forall i \in \VV$.
%{\color{blue}For some applications, it is important to have the individual rationality. Then, the aforementioned constraint should be used in the bi-level optimization problem.}
%Otherwise, non-participating players may receive a negative share of the lottery and pay for this lottery even if they rejected to participate in it.
\end{remark}

\section{Case study}\label{sec:case}
We apply our perturbed lottery to demand response in the smart grid. Demand response involves a load serving entity (LSE) and a set of end-users. The LSE is the social planner and wants to incentivize the end-users to shift their peak-time demand to off-peak time. 
The end-users participate in the lottery by shifting a portion of their shiftable demands.

Consider a power transmission network $(\GG,\EE)$ where $\GG$ and $\EE$ denote the set of buses and the set of transmission lines, respectively. In particular, $\VV \subseteq \GG$, and $\PP \subseteq \GG$ denote the set of load buses with non-zero demand, and the set of generator buses, respectively. Each line $l \in \EE$ has power flow capacity $f^{\max}_l \in {\real}_{\geq 0}$ and $f^{\max}=[f^{\max}_1,\cdots,$ $f^{\max}_{|\EE|}]^T$.

\begin{figure}
  \centering
  \includegraphics[width = \linewidth]{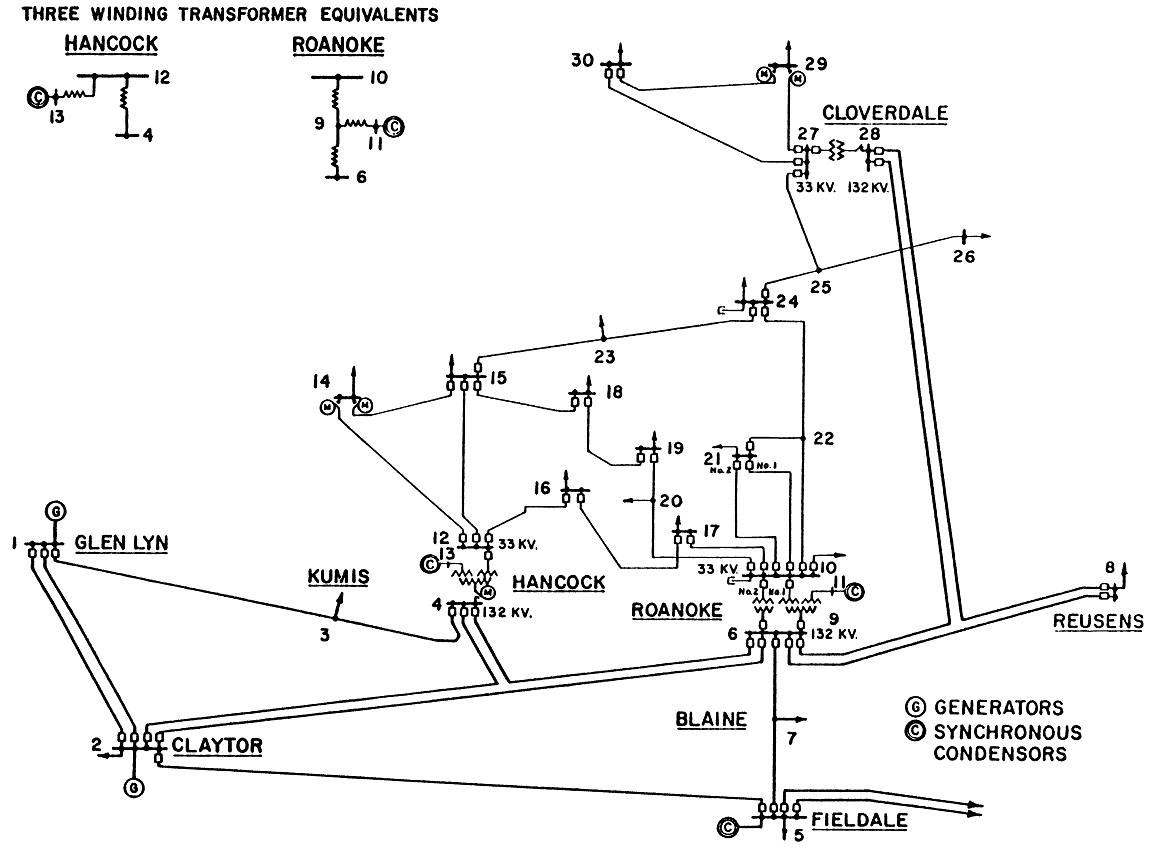}
  \caption{IEEE 30-bus test system~\cite{christie2000power}.}\label{30bus600}
\end{figure}

With the perturbed lottery, each end-user has payoff function~\eqref{e14}
where decision variable $s_i$ denotes shifted demand in monetary value. 
Function $h_i$ represents any impetus from the public good; e.g., inverse of stability concern, utility discount, additional rewards made by the LSE.
The LSE solves problem~\eqref{e17}, in which convex constraints $g$ represent three physical constraints; i.e.,
the end-users cannot shift more than the demand, and the total adjusted demand after shifting cannot exceed the total power generation, and the line capacities are enforced:
\begin{align}
&L-s^* \geq 0, \ \sum_{i \in \VV}(L_i-s_i^*) \leq \sum_{j \in \PP}P_j, \nnum\\
&-f^{\max} \leq H_p P-H_l(L-s^*) \leq f^{\max}
\label{mm05}
\end{align}
where $L \in {\real}_{\geq 0}^{|\VV|}$ and $P \in {\real}_{\geq 0}^{|\PP|}$ denote power demand and power generation, respectively.
Matrix $H \in [-1,1]^{|\EE| \times |\GG|}$ is the injection shift factor matrix where $(a,b)$ entry of $H$ represents the active power change on line $a$ with respect to change in power injection at bus $b$. In particular, matrices $H_l \in [-1,1]^{|\EE| \times |\VV|}$ and $H_p \in [-1,1]^{|\EE| \times |\PP|}$ are the collections of columns $i \in \VV$ and $i \in \PP$ of $H$, respectively.
Since $L$, $P$, $f^{\max}$ are constants at the given time, constraints~\eqref{mm05} are convex and thus satisfy Assumption~\ref{nasm3}.
It is worth noticing that the physical interconnections~\eqref{mm05} are captured by the constraints $g(s^*(R),R)\leq0$ in the social planner's problem~\eqref{eq009}, not by the payoff model.

%We conduct case studies using IEEE 30-bus test system where $|\PP|=6$, $|\VV|=20$, and $|\EE|=41$. Topology of IEEE 30-bus test system can be found in~\cite{christie2000power} where each bus $i \in \GG$ is associated with the bus number.

We conduct case studies using IEEE 30-bus test system shown in Figure~\ref{30bus600} where $|\PP|=6$, $|\VV|=20$, and $|\EE|=41$.
The system parameters are obtained from MATPOWER~\cite{zimmerman2011matpower}. Money/power exchange rate $\$0.1 / kWh$ is applied and $1$ hour time frame is considered; e.g., the generator at bus $1$ generates $23.54 MW \times 1 h \times \$0.1 / kWh = \$2354$.
Each load bus's power demand increases $30\%$ without changing power generations, so demand shifts are inevitable.
%We intentionally increase the power demand of each load bus by $30\%$ without changing power generations, so that demand shifts are inevitable.

%\begin{figure}   \centering   \includegraphics[width = \linewidth]{Figures/30bus600}   \caption{\color{blue}Topology of IEEE 30-bus test system~\cite{christie2000power} where   each bus $i \in \GG$ is associated with the bus number. Of them, 20 buses are load buses (i.e. 20 players).}\vspace{-0.5cm}\label{30bus600} \end{figure}

We choose $h_i(\bar{s}-R) = (100+i) \log(\bar{s}-R+1)$ for bus $i \in \VV$; e.g., bus $30 \in \VV$ has $h_{30}(\bar{s}-R) = 130 \log(\bar{s}-R+1)$.
One can see that
function $h_i$ satisfies Assumption~\ref{nasm1}.
%One can see that function $h_i$ satisfies Assumption~\ref{nasm1}.
The logarithmic model of provision of public good $h_i$ is based on Cobb-Douglas utility function~\cite{cobb1928theory}.
Recent papers~\cite{mccleary2006us,ribar2002altruistic,rotemberg2014charitable} use such function to express the benefits from a public good. We choose $\alpha=1$.

The socially optimal public good $G^*=\$2317$ of the unperturbed lottery is calculated by~\eqref{er01}.
The socially optimal payoff is obtained by $\sum_{i \in \VV}h_i(G^*)-G^*=\$7142$.

%The LSE solves problem~\eqref{e17}, in which convex constraint $g(s^*(R,c),R,c)=[s_{\min} -s_1^*,\cdots,s_{\min} -s_N^*,\bar{s}_{\min} -\bar{s}]^T \leq \vec{0}^{\,}$ represents the desired amount of minimal demand shifts with non-negative constants $s_{\min}$, and $\bar{s}_{\min}$. We choose $s_{\min}=\$ 3$, $\bar{s}_{\min}=\$ 350$, and $\alpha=1$.

\begin{figure}
  \centering
  \includegraphics[width = \linewidth]{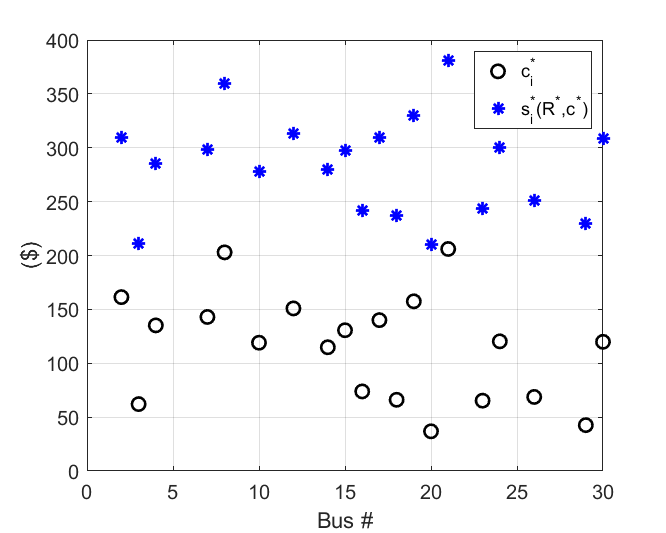}
  \caption{Optimal solution $c_i^*$ with $R^*=\$3358$ and the corresponding Nash equilibrium $s_i^*$. The bus numbering (horizontal ordinate) can be found in Figure~\ref{30bus600}.}\label{Simul_1}
\end{figure}

\begin{figure}
  \centering
  \includegraphics[width = \linewidth]{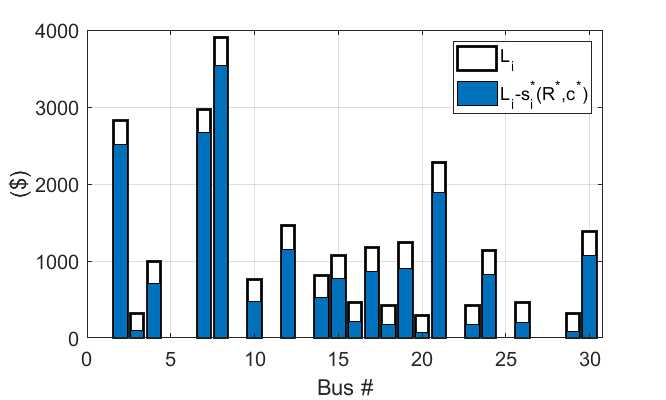}
  \caption{Power demand $L_i$ and adjusted demand $L_i-s_i^*(R^*,c^*)$ after shifting.}\label{Simul_30_3}
\end{figure}

\begin{figure}
  \centering
  \includegraphics[width = \linewidth]{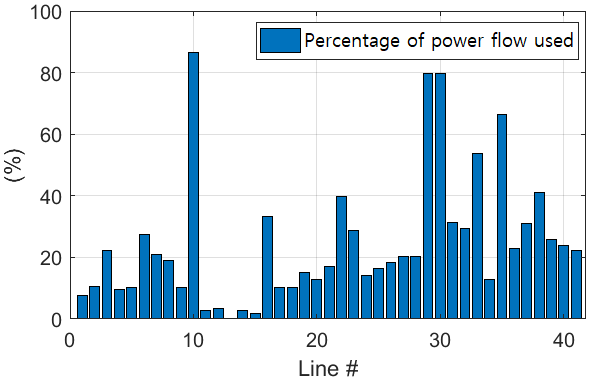}
  \caption{Percentage of power flow used in each line. The line numbering (horizontal ordinate) can be found in~\cite{christie2000power}.}\label{Simul_30_const1}
\end{figure}

We solve problem~\eqref{e17} by CVX~\cite{grant2008cvx}, and generate optimal value $\$5675$ with solution $(R^*,c^*)$ presented in Figure~\ref{Simul_1}. The figure also presents the induced Nash equilibrium of the optimal lottery game.
The aggregate payoff induces the socially optimal public good $\bar{s}^*(R^*,c^*)-R^*=\$5675-\$3358=\$2317=G^*$, and the socially optimal payoff $\sum_{i \in \VV}h_i(G^*)-G^*=\$15644$.
Convex program~\eqref{new.e0} generates a large reward $R^*=\$3358$ to satisfy the physical constraints~\eqref{mm05}.
Note that $\bar{c}=G^*$ is a sufficient and necessary condition for the optimality, according to Theorem~\ref{the51}.
%In the solution, $c_i$ is in a descending order based in their indices. This is because players who have small gain $(1+\frac{i}{100})$ for the marginal benefit function should be stimulated to invest to satisfy constraint $s_i^*(R,c)\geq s_{\min}$.
By Theorem~\ref{the51}, the solution is identical to that of problem~\eqref{e17} and satisfies all the physical constraints described in~\eqref{mm05}. Figure~\ref{Simul_30_3} visualizes that the first constraint is satisfied where the shifted demand never exceeds the power demand. The second constraint is also satisfied because $\sum_{j \in \PP}P_j = \sum_{i \in \VV}(L_i-s_i^*)=\$18921$.
Figure~\ref{Simul_30_const1} shows that power flow at each transmission line never exceeds its capacity.
As shown in this simulation, the optimal bi-level lottery provides the optimal solution that maximizes the aggregate utility, while guaranteeing convex constraints imposed by the social planner.


\section{Conclusions} This paper studies an optimal bi-level lottery design problem where a social planner aims to achieve the social optimum through the least reward and perturbations. We approximate the problem via a convex relaxation and identify mild sufficient conditions under which the approximation is exact. The results are verified via a case study on demand response in the smart grid.

\bibliographystyle{unsrt}

\end{document}